\documentclass[conference]{IEEEtran}

\usepackage{cite}

\ifCLASSINFOpdf
 \usepackage[pdftex]{graphicx}

\usepackage{amssymb}
\usepackage{lineno}
\usepackage{lscape}
\usepackage{tabularx}
\usepackage[gen]{eurosym}
\usepackage{multirow}
\usepackage{hhline}
\usepackage{booktabs}
\usepackage{siunitx}
\usepackage{float}
\usepackage{framed}

\else

\fi

\begin{document}
%
\title{
Social Influence in Agile Requirements Engineering}



\author{\IEEEauthorblockN{Lucas Gren}
\IEEEauthorblockA{Chalmers University of Technology and The University of Gothenburg\\
Gothenburg, Sweden 412--92\\
Email: lucas.gren@cse.gu.se}

}

\maketitle

\begin{abstract}
Agile requirements engineering implies more complex communication patterns since even the developers are supposed to have direct contact with customers. With more face-to-face communication comes social-psychological factors influencing the requirements. Studies have pointed at the importance of negotiation training, but I argue that more basic human traits can be triggered in favor of the negotiator with the most knowledge of social influence research and practice. I suggest a plan of how research in social influence and requirements facilitation can be conducted, mostly through experimentation. 
\end{abstract}

\IEEEpeerreviewmaketitle
\section{Introduction}
In a recent systematic literature review, \cite{inayat2015systematic} concluded that studies on agile requirements engineering are still rare and more studies are needed in order to better understand the practices involved. The paper still identifies some Agile RE challenges, e.g.\ customer availability, neglect of non-functional requirements, customer inability and agreement, to name a few. Customer inability and agreement is an interesting agile RE challenge since the customer are supposed to have direct communication with the developers, and is preferably even on site \cite{ramesh2010agile}. The authors highlight that studies are needed especially in connection to non-functional requirements and self-organizing teams. Since then, at least one study on agile teams have provided additional explanation to the agile team dynamic from a psychological perspective \cite{grenjss2}. However, there are more theories from social psychology that might be useful in order to understand how agile teams negotiate with customers since there is obviously more communicative complexity in such agile practices.

\section{Negotiation and Social Influence in Agile Requirements Engineering}
\begin{figure}
\centerline{\includegraphics[scale=0.55]{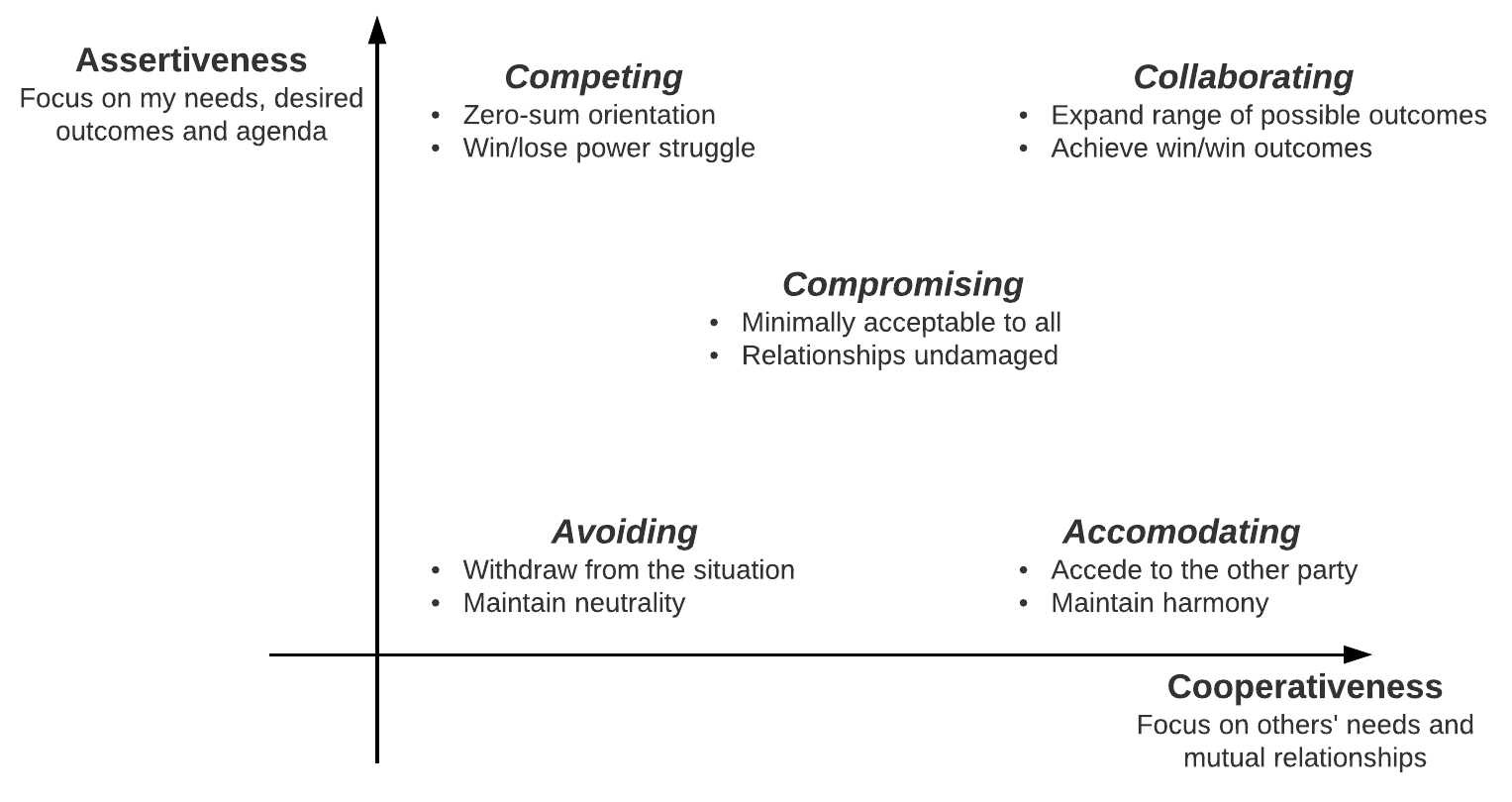}}
\caption{The Thomas-Kilmann Conflict Modes (adopted from \cite{thomas1992conflict}).}
\label{kilmann}
\end{figure}

Some researchers have stressed the importance of teaching software engineers negotiation skills, since an extensive body of knowledge already exists in the field of organizational psychology (see e.g.\ \cite{ahmad2011empirical}). In \cite{ahmad2008negotiation}, the author suggests the use of the famous Tomas-Kilmann Conflict Modes. It is important to recognize that different people have different interests in different negotiations. A well-used model of such stance in negotiation is the Tomas-Kilmann Conflict Modes \cite{thomas1992conflict} and is shown in Figure~\ref{kilmann}. Depending on the assertiveness and cooperativeness in each situation a person will approach the conflict mainly in five different ways (although people have a tendency to resort to some of them more than others). 

In the software requirements engineering context, Kovitz \cite{kovitz2003hidden} states that: ``In practice, the requirements phase is really a negotiation phase. The customers and developers try to find a set of features that they can implement within a time that delivers acceptable return on investment for the customer. Especially in phased development, the developers can feel a great deal of pressure to make promises that they can't keep.''

Knowing the negotiation styles in context might help the developers to understand what happens in these contexts. However, and just like Kovitz \cite{kovitz2003hidden} continues: ``So, during the requirements phase, the requirements engineer often finds herself or himself in the role of negotiator, trying to find a realistic schedule and present it in a way that is acceptable to the customer. Many requirements engineers are familiar only with the technical side of software development and are not prepared for the kind of gamesmanship that occurs in negotiation with a savvy customer. Yet this negotiation may be the single most important factor in the success of the project. In larger organizations, politics often overrides technical factors, and phased development puts the requirements engineer into the thick of politics –-- whether (s)he knows it or not.''

Not only does organizational politics play a proven key role (see e.g.\ \cite{buchanan2008you}), but more basic cognitive traits can easily be used in negotiation to gain advantages \cite{cialdini2004social}. I aim at conducting empirical research to observe and analyze the effects of such compliance techniques (compliance refers to a particular kind of response, that of acquiescence, to a particular kind of communication, that is, a request \cite{cialdini2004social}), taken from social influence research. We will now present some of the techniques we plan on studying in the requirements engineering context.

\section{Social Influence}
All people are motivated to achieve their goals in an effective and rewarding way. People have a desire to respond appropriately to a social situation, which requires an accurate perception of reality. In such compliance-gaining attempts, it is of utter importance to interpret and react to information correctly. There is s set of states and techniques that have been proven to increase compliance \cite{cialdini2004social}, some of which are presented next together with possible connections to agile requirements engineering. 

\subsection{Affect and Arousal}
Affect and emotional state have been shown by many researchers to have an effect on compliance (see e.g.\ \cite{forgas1995mood}). Fore example, \cite{dolinski1998fear} showed that people have significantly more compliance right after a strong emotion of fear. They call this effect Fear-Then-Relief and show that we should be extra alert ``when the danger is over.'' In an RE context, it would be interesting to see if such triggers of fear (like stating a severe problem with the developed system) followed by an incidental removal of fear would affect negotiators to show increased compliance.

\subsection{That's-Not-All Technique}
Another effective technique is to make use of the fact that people often do not have time to thoroughly assess the situation and use entirely and deliberate evidence-based decision-making \cite{cialdini2004social}. Burger \cite{burger1986increasing} describes the technique of That's-Not-All consisting of making an initial request followed by an immediate better deal by reducing costs or increased benefits of compliance. The effect can be explained by reciprocation, but also by using the contrast between the two options and the shift of anchor points, meaning that the customer then expands the acceptable cost interval, for example \cite{burger1986increasing}. It is easy to see how such initial overpricing is used in RE today, but we would like to investigate the extent of these techniques empirically through experimentation.

\subsection{Foot-In-The-Door Technique}
Yet another technique is to use the basic human trait of consistency. People want to be consistent in their behavior, statements, beliefs, and self-ascribed traits. The idea is to make a very small request initially, that the person almost has to answer affirmatively, and then make a larger, but often related, request. Such a technique has been shown effective in many studies over the year, but initially by \cite{freedman1966compliance}. In order to run an Agile RE experiment on the effect, one could test providing an initial small request in one group and not in the control group, essentially replicating the studies conducted by \cite{freedman1966compliance}, but in an RE context. This would be interesting since we do not know if or how a requirements negotiation context differs. 


\section{Discussion}\label{sec:disc}
While I have stated that more negotiations are happening in agile requirements engineering, some of these proposed studies would be applicable to more requirements engineering approaches. Replicating these social influence studies in an RE context would surely increase the understanding of the requirements elicitation process and maybe trigger a will for practitioners to educate their negotiators (especially in agile RE since many more often negotiate using such a process). One would expect some of the effects to differ in an RE context since stakeholders are often more personally engaged in eliciting requirements for a software system. However, under high cognitive load, unwanted decisions might be more likely, which then could be avoided with more knowledge on how social influence affects compliance in the RE context. 

Maybe we do not have the opportunity to train all the developers in negotiation and social influence practices at once, and in such cases, I believe having key expert negotiators that facilitates negotiations between the developers and the customer (much like the process facilitator role of the Scrum Master) in order to control unwanted influence of decisions, would be a vast improvement still. However, all these claims are to be investigated empirically, and eventually an output from our research will be the development of a toolbox and best practices on how to conduct agile requirements engineering in a structured and well-planned manner based on social influence and negotiation research.

\bibliographystyle{IEEEtran}

\bibliography{references}

\end{document}